\documentclass[reprint,aps,prl,twocolumn,superscriptaddress]{revtex4-1}
\usepackage{braket}
\usepackage{amssymb}
\usepackage{amsmath}
\usepackage{epsfig}
\usepackage{color}
\usepackage{graphics, graphicx}
\usepackage{bbold}
\usepackage{psfrag}
\usepackage{mathcomp}
\usepackage{subfigure}
\usepackage{verbatim}
\usepackage[colorlinks, citecolor=blue]{hyperref}
\usepackage[normalem]{ulem}
\usepackage{bm}
\def\cp#1{\mathbf{#1}}
\begin{document}

\title{Quantum diatomic chain: a supersolid structure in three-component Bose mixture}
\author{Francesco Ancilotto}
\affiliation{Dipartimento di Fisica e Astronomia ``Galileo Galilei''
and CNISM, Universit\`a di Padova, via Marzolo 8, 35122 Padova, Italy}
\affiliation{ CNR-Istituto Officina dei Materiali (IOM), via Bonomea, 265 - 34136 Trieste, Italy }

\date{\today} 

\begin{abstract}

The formation and properties of a supersolid 
structure in a three-component ultracold Bose gas mixture at T=0
are investigated theoretically. The system consists of $^{23}$Na, $^{39}$K, 
and $^{41}$K atomic species, in which the binary mixtures of ($^{23}$Na,$^{39}$K) and ($^{39}$K,$^{41}$K) 
can form self-bound quantum droplets stabilized by quantum fluctuations. 
Two such droplets can bind together by the shared $^{39}$K component, forming
a stable "dimer" structure, which 
displays vibrational modes analogous to a classical diatomic molecule. 
A simple protocol is proposed to create a stable linear
chain formed by periodic repetition of this basic building block, i.e.
an alternating sequence of ($^{23}$Na,$^{39}$K) and ($^{39}$K,$^{41}$K) droplets. 
This structure exhibits both periodic density modulations from the 
droplet ordering and global phase coherence due to the shared $^{39}$K component, 
satisfying the criteria for supersolidity.
This expands the class of known supersolids by adding a
system where mediated binding — rather than intrinsic long-range interactions
or engineered band-structures as in previously known supersolids — is the key organizing principle,
thereby offering new directions for both theory and experiment.
The low-energy excitation spectrum, probed by density perturbations, 
identifies modes corresponding to droplet vibrations
close to the ones expected from a classical diatomic chain,
coexisting with low-energy superfluid (Goldstone-type) modes.

\end{abstract}
\maketitle

\section{I. Introduction}

The study of ultracold quantum gases has been at the forefront of atomic, 
molecular, and optical physics for decades now, 
often unveiling new states of matter and quantum phenomena \cite{bloch_08,chin,baroni,choma}. 
Among the fascinating phenomena observed in these systems, quantum droplets and supersolids 
have gained significant attention due to their unique properties and 
potential for advancing our understanding of quantum matter.

Quantum droplets were first predicted theoretically by Petrov in 2015 \cite{petrov_15} 
and subsequently observed experimentally 
in dipolar gases \cite{kadau,ferrier,Schmitt2016,Ferlaino2016} 
as well as in homonuclear \cite{Cabrera2018,Semeghini2018} and heteronuclear 
\cite{Derrico2019,Guo2021} mixtures.
These self-bound states are stabilized by quantum fluctuations, 
typically described by the Lee-Huang-Yang (LHY) correction \cite{lhy}, 
which counteracts the mean-field attraction that would otherwise 
lead to collapse. 
Quantum droplets are self-bound states characterized 
by ultralow equilibrium densities and surface tension.
These droplets exhibit liquid-like properties while maintaining quantum coherence, 
making them a unique form of matter that bridges the gap between classical liquids and quantum gases.

Supersolidity, a paradoxical state of matter combining the seemingly 
contradictory properties of crystalline order and frictionless flow, 
was first proposed by Andreev and Lifshitz in 1969 \cite{andreev} and
Leggett in 1970 \cite{leggett} (see also Ref.\cite{gross} for an earlier proposal of
a solid phase of interacting Bosons). After decades of inconclusive searches in 
helium systems \cite{chan}, the first clear demonstrations of 
supersolidity were achieved in ultracold atomic gases, 
specifically in dipolar quantum gases \cite{tanzi,bottcher,chomaz},
spin-orbit-coupled Bose-Einstein condensates \cite{li},
and by coupling BEC in optical cavities \cite{leonard}.
Theoretical results for Rydberg-dressed atoms and soft core bosons suggested
the possibility of realizing supersolid structures also in these systems \cite{
homeier,cinti_10,sacca,sac12,anci_13,rev1,rev2}.
Recent theory predicts the existence of binary dipolar
supersolids, where two dipolar superfluids combine to form
a periodically modulated state \cite{blan22,li1}.

The possibility of supersolid phases in a binary Bose mixture
has recently been put forward in Ref. \cite{reimann,Sac20}, where a self-bound 2D
supersolid stripe-phase in a weakly interacting binary BEC with spin-orbit coupling
has been proposed, being stabilized by the Lee-Huang-Yang beyond-mean-field term.
In Ref. \cite{Ten22}, a single one-dimensional droplet made
of a binary Bose mixture immersed into a background of the excess species
and subject to periodic boundary conditions (as a model for a droplet
confined in a toroidal trap), was found to display
non-classical rotational inertia and thus the coexistence
of rigid-body and superfluid character.

It has been suggested \cite{anci23} that a droplet array resulting from
the breakup, caused by the Rayleigh-Plateau instability, of a 
quantum liquid filament made
of a K-Rb Bose mixture, and where atoms of one species are
in excess with respect to the optimal equilibrium ratio, may 
display supersolid character.

The rich physics of quantum droplets and supersolids has primarily been 
investigated in single- and two- components dipolar gases and two-component Bose mixtures,
whereas the possibility of observing these phenomena in three-component systems remains 
largely unexplored.
In a recent theoretical work \cite{borromean} cluster formation in three-component ultracold bosons 
has been investigated, in the form of Borromean cluster, where only the ternary 
bosons can form a self-bound droplet while any binary subsystems cannot.
A paper appeared later\cite{ma} where a new type of shell-shaped
Bose-Einstein condensate with a self-bound character has been proposed, made of three-component 
$^{23}$Na $^{39}$K $^{41}$K 
Bose mixture (species (1,2,3) in the following),
where the mixtures (1,2) and (2,3) both form self-bound droplets. In the proposed
structures an outer shell of liquid (1,2) envelops a spherical 
core made of the (2,3) liquid, and it was claimed to be stable without the need of any trapping potential.
As shown in Ref.\cite{comment}, however, it turns out that 
these structures are not actually the ground-states
solutions to the system but rather metastable states corresponding to 
local energy minima. The lowest energy state is instead a "dimer" configuration where
two quantum droplets (made of the (1,2) fluid and (2,3) fluid, respectively)
are bound together by the shared component 2 \cite{comment}.

Starting from this finding, 
I found that a simple protocol (easily implementable in real experiments)
may results in the formation of a more complex structure where multiple dimers are 
bound together in a linear structure, which has supersolid character.

The system described here consists of linear chains of 
alternating quantum droplets (quantum "diatomic" linear chain), each 
composed of mixtures of components (1,2) and (2,3), where 
component (2) acts as a "glue" connecting adjacent droplets, and it is stable in vacuum, i.e. no 
trapping potential is required for its stabilization.
This structure displays supersolid properties, in the form of finite non-classical 
translational inertia.
This configuration presents therefore a unique opportunity to study the 
interplay between quantum droplet formation, supersolidity, 
and the role of a mediating component in a reduced-dimensional setting.

The remainder of this paper is organized as follows: In Section II, 
the theoretical model and the numerical methods used for the
calculations are presented. 
Section III describes results for the building block of the proposed 
supersolid, i.e. the quantum dimer. In Section IV the properties of the
"diatomic" linear chain made by periodic repetitions of dimers along the chain axis,
and its supersolid character are investigated, together with its vibrational properties.
Finally, Section V contains a summary and outlook for future research directions.

\section{II. Method}

The system under study here is the same as in Ref.\cite{ma}, 
i.e. a three-component $^{23}$Na $^{39}$K $^{41}$K Bose mixture,
at zero temperature and in the absence of three-body recombination effects.
An inhomogeneous mixture made of the above species is described within 
the density functional theory (DFT) approach in the MF+LHY framework, where
the total energy functional is given by (atomic units will be used hereafter)

\begin{widetext}
\begin{equation}
E = \sum_{i=1}^{3}\int d\bm{r} \, \frac{1}{2m_i}|\nabla \psi_{i}(\bm{r})|^2  
+\frac{1}{2}\sum_{i,j=1}^{3}g_{ij}\int d\bm{r} \, \rho_{i}(\bm{r})\rho_{j}(\bm{r}) 
+\int d\bm{r} \, {\cal E} _{LHY}(\rho_1(\bm{r}),\rho_2(\bm{r}),\rho_3(\bm{r})) 
\label{eq:energy2c}
\end{equation}
\end{widetext}

Here $\rho_i(\bm{r})=|\psi _i(\bm{r})|^2$ 
represent the number density of each component
($i=1$ for $^{23}$Na, $i=2$ for $^{39}$K and $i=3$ for $^{41}$K). 
The coupling constants between species $i$ and $j$ are 
$g_{ij}=2\pi a_{ij}/m_{ij}$,
with scattering length $a_{ij}$ and reduced mass $m_{ij}=m_im_j/(m_i+m_j)$. 
The number densities $\rho_i$ are normalized such
that $\int _V \rho_i(\bm{r})\,{\rm d}\bm{r} =N_i$ $(i=1,2,3)$, where $N_i$ are the 
total number of atoms in the $i$-th component.

Components 1 and 3 interact via strong repulsive potential and are
therefore immiscible, while the binary mixtures (1,2) and (2,3) 
separately form self-bound binary droplets. As shown in Ref.\cite{ma}, this 
is achieved with 
$(a_{11},a_{22},a_{33},a_{12},a_{13})=(52, 30, 63, -50, 213)a_0$ ($a_0$ is the Bohr radius),
while $a_{23}$ (which is highly tunable via a Feshbach resonance) must have a 
negative, and sufficiently large, value.
I will consider here the same two representative values used in Ref.\cite{ma}, 
$a_{23}=-70\,a_0$ and $a_{23}=-200\,a_0$,
although, due to the heavy computational cost,
most of the calculations will be performed with $a_{23}=-200\,a_0$.
The choice of this value is motivated by the fact that it gives 
stronger binding between droplets and therefore clearer signals 
in the computed spectra (as discussed in the following Sections).

The term accounting for quantum fluctuations is 
\begin{equation}
{\cal E} _{LHY}=\int \frac{d^3{\bf k}}{2(2\pi)^3}\left[\sum_i(E_{i{\bf k}}-
\epsilon_{i{ k}}-g_{ii}\rho _i)+\sum_{ij}\frac{2m_{ij}g_{ij}^2\rho _i \rho _j}{{\bf k}^2}\right] \label{E_qf}
\end{equation}  
with $\epsilon_{i{ k}}=k^2/2m_{i}$, and $E_{i{\bf k}}$ is the $i$-th Bogoliubov excitation energy
\cite{borromean}. 
The three dispersion relations $E_{i{\cp k}} \ (i=1,2,3)$ 
are the roots of the following equation \cite{borromean}
\begin{equation}
x^3+bx^2+cx+d=0,
\end{equation}
where
\begin{widetext}
\begin{eqnarray}
b&=&-\sum_i \omega_{i}^2,\\
c&=&\sum_{i<j}\left((\omega_{i}\omega_{j})^2-4g_{ij}^2\rho _i \rho _j\epsilon_{i{ k}}\epsilon_{j{ k}}\right),\\
d&=&-(\omega_{1}\omega_{2}\omega_{3})^2-16\epsilon_{1{ k}}\epsilon_{2{ k}}\epsilon_{3{ k}}
g_{12}g_{23}g_{13}\rho_1 \rho_2 \rho _3+\sum_{i<j, l\neq(i,j)}4\epsilon_{i{ k}}
\epsilon_{j{ k}}\rho _i \rho _j g_{ij}^2\omega_{l}^2.
\end{eqnarray}
\end{widetext}
Here $\omega_{i}=\sqrt{\epsilon_{i{ k}}^2+2g_{ii}\rho _i\epsilon_{i{ k}}}$ ($i=1,2,3$) 
are the Bogoliubov spectra for the individual components \cite{note}. 

Minimization of the action associated to Eq.\eqref{eq:energy2c} leads
to the following Euler-Lagrange equations
\begin{equation}
\begin{split}
i\frac{\partial \psi_i (\vec r, t)}{\partial t}=\left(-\frac{\nabla^2}{2m_i}+\sum_{j}g_{ij}\rho _j+\frac{\partial 
{\cal E}_{LHY}}{\partial \rho_i}\right)\psi_i (\vec r, t)
\end{split} \label{GP}
\end{equation}

The numerical solutions of Eqs.\eqref{GP} provide
the real-time evolution of the system in three-dimensional space.
When stationary states are sought the left hand side of
Eq.\eqref{GP} is replaced by $\mu _i\psi_i (\vec r)$,
where $\mu _i$ is the chemical potential of the $i$-th species. 
The evolution in imaginary time (via, e.g., steepest descent algorithm)
allows to obtain stationary state solutions starting 
from suitable initial wavefunctions. The chemical potentials 
$\mu _i$ are determined iteratively so
that the target values of $N_i$ are achieved. 

The wave functions $\{\psi_i \}$ are mapped on an
equally spaced 3D Cartesian grid and the Laplacian operator in Eq.\eqref{GP}
is represented by a 13-point finite-difference stencil.
The equations \eqref{GP} are propagated in real time
by using the Hamming's predictor-modifier-corrector method
initiated by a fourth-order
Runge-Kutta-Gill algorithm \cite{Ral60}.
Periodic boundary conditions (PBC) are imposed along the three spatial directions.
The mesh spacing and time step are chosen such that
during the real time evolution excellent conservation
of the total energy of the system is guaranteed.

\section{III. The "dimer" configuration}

The self-bound, shell-shaped structure investigated in Ref.\cite{ma}
is made of three-component $^{23}$Na $^{39}$K $^{41}$K 
Bose mixture,
where the mixtures (1,2) and (2,3) both form quantum droplets. In the structure proposed
in Ref.\cite{ma} an outer shell of liquid (1,2) envelops a spherical 
core made of the (2,3) liquid, and it is claimed to be stable without the need of any trapping potential.
However, this turns out to be a {\it metastable} state \cite{comment}:
any perturbation would make these structures swiftly
convert into the actual lowest energy state, characterized by 
a "dimer" configuration, where
two self-bound droplets (made of the (1,2) fluid and (2,3) fluid, respectively)
are kept in mutual contact by the shared component 2.

\begin{figure}[t]
\includegraphics[width=7cm]{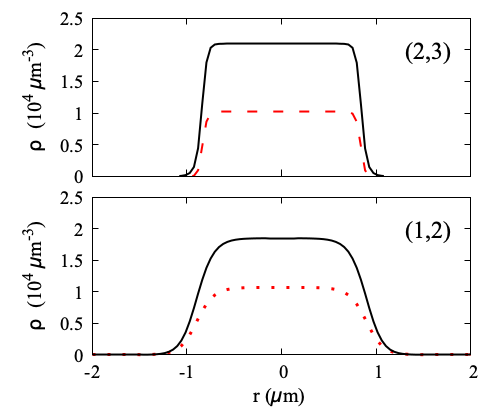}
\caption{Density profiles of the individual self-bound droplets made of
liquid (2,3) (upper panel) and (1,2) (lower panel). 
Dotted line: species $1$; solid line: species $2$; dashed line: species $3$.  
} 
\label{fig1}
\end{figure}

In Fig.\ref{fig1} the
equilibrium densities of two separate self-bound droplets are shown,
for the case $a_{23}=-200\,a_0$.
The first (1,2) droplet is made of $N_1=3.5\times 10^4$ and $N_2=5.9\times 10^4$ atoms, while the second 
(2,3) droplet is
made of $N_2=5.4\times 10^4$ and $N_3=2.5\times 10^4$ atoms.
The total masses of each droplet, which will be relevant in the discussion of the 
dimer vibrational properties, are $M_{12}=N_1m_1+N_2m_2=5.66\times 10^9$ and
$M_{23}=N_2m_2+N_3m_3=5.71\times 10^9$ (expressed in atomic units).

In the interior of each droplet, the equilibrium density 
ratio $\rho _j/\rho _i$ is locked at the value $\sqrt{g_{ii}/g_{jj}}$ \cite{petrov_15,ma},
as expected from the theory of binary Bose mixtures.

\begin{figure}[t]
\includegraphics[width=7cm]{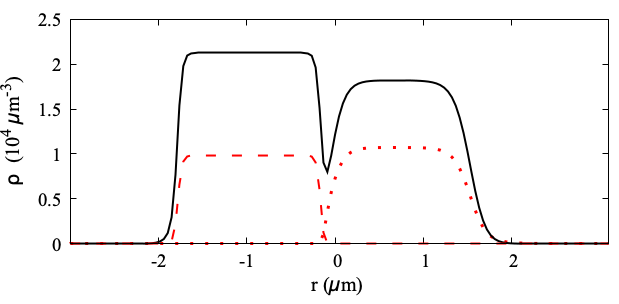}
\caption{Dimer density profile along the line passing through its axis,
for the case $a_{23}=-200\,a_0$.
Dotted line: species $1$; solid line: species $2$; dashed line: species $3$.}

\label{fig2}
\end{figure}

\begin{figure}[t]
\includegraphics[width=7cm]{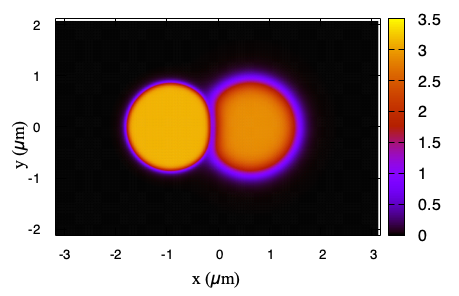}
\caption{Total density map corresponding to the dimer
structure shown in Fig.\ref{fig2}. 
The droplet (2,3) is on the left, the droplet (1,2) on the right.
The density units are the same as in Fig.\ref{fig2})
} 
\label{fig3}
\end{figure}

In Fig.\ref{fig2} and Fig.\ref{fig3} is shown the "dimer" equilibrium structure
resulting from the interaction between the two droplets, 
the binding being a consequence of sharing the component (2) between the two droplets.
Notice that, while the number of atoms $N_1$ and $N_3$ can be fixed independently,
the number of atoms $N_2$ should be taken close to the ideal 
number $N_2=N_2^{(1)}+N_2^{(2)}$, where $N_2^{(1)}$ and $N_2^{(2)}$ are such 
to satisfy the equilibrium ratio in the interior of the droplets, 
$N_1/N_2^{(1)}\sim \sqrt{g_{22}/g_{11}}$ and 
$N_2^{(2)}/N_3\sim \sqrt{g_{33}/g_{22}}$.

\begin{figure}[t]
\includegraphics[width=7cm]{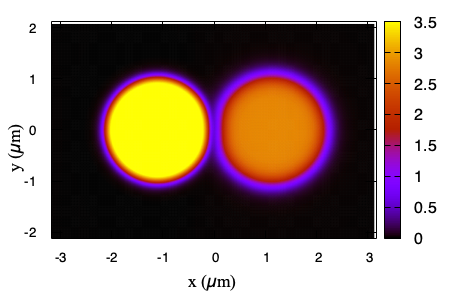}
\caption{Total density map 
corresponding to the dimer structure with $a_{23}=-70\,a_0$.
The droplet (2,3) is on the left, the droplet (1,2) on the right.
The density units are the same as in Fig.\ref{fig2}.
} 
\label{fig4}
\end{figure}

For comparison, the stable dimer configuration is shown in Fig.\ref{fig4}
for a lower value $a_{23}=-70\,a_0$.
It appears that due to the weaker coupling between species (2,3), 
the contact area between the two droplets is reduced
with respect to the case $a_{23}=-200\,a_0$.
The resulting lower binding between the two droplets will show up, as discussed in the
following, in a lower frequency of the dimer bond-length oscillations.

Below, I briefly discuss the vibrational properties
of the dimer structure shown in Figs.\ref{fig2}-\ref{fig3}.

It is useful in this regard to preliminary study 
the intrinsic oscillations of the individual droplet components (see Fig.\ref{fig1}).
In particular, the frequencies $\omega _{l,n}$ of the lowest 
angular momentum modes will be computed, i.e. $l=0,n=0$ (monopole "breathing" mode)
and $l=2,n=0$ (quadrupole mode). The frequency of the $l=1,n=0$ (dipole mode) is zero,
corresponding to a rigid translation of the droplet as a whole.

To extract the frequency of such modes, I applied to the ground-state 
wavefunctions $\psi _i$ of each droplet  
(providing the equilibrium configurations shown in Fig.\ref{fig1})
a perturbation such that the initial states for the dynamics 
are $\psi_i ^{ini}=e^{i\epsilon F}\psi _i$
where $F=r^2$ for the monopole mode, $F=(3z^2-r^2)$ for the
quadrupole mode, and $\epsilon $ is a small number. 
The system is then allowed to evolve in real-time starting from $\psi_i ^{ini}$.
The average value of the moments during the dynamics is recorded,
$<F(t)>=<\psi _i(t)|F|\psi _i(t)>$, and finally it is Fourier transformed
into the frequency domain, $\tilde {F}(\omega )=\int <F(t)>e^{i\omega t}dt$.
The peaks in $\tilde {F}(\omega )$ allow to extract the sought frequencies.
I found $\omega ^{(1,2)}_{quad}=1198$ Hz, while $\omega ^{(2,3)}_{quad}=2045$ Hz.
The monopole modes have much
higher frequencies, $\omega ^{(1,2)}_{mono}= 4850$ Hz and $\omega ^{(2,3)}_{mono}= 17700 $ Hz.



\begin{figure}[t]
\includegraphics[width=7cm]{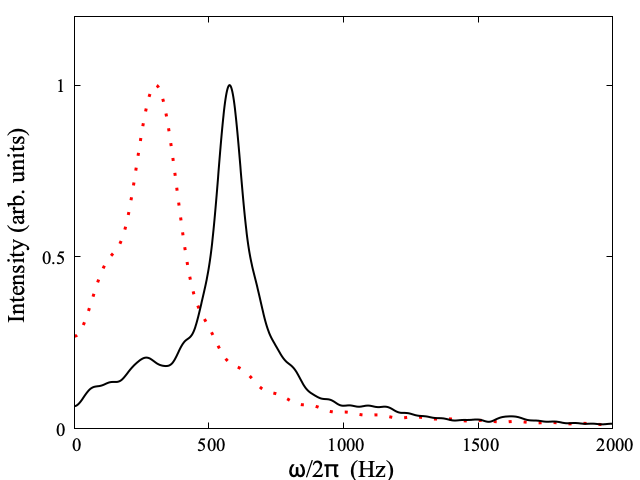}
\caption{Spectrum for the dimer oscillations. Solid line: $a_{23}=-200\,a_0$, dotted line: $a_{23}=-70\,a_0$. 
} 
\label{fig5}
\end{figure}

Interestingly, the dimers shown in Fig.\ref{fig3} and Fig.\ref{fig4} display simple vibrational patterns 
similar to a classical spring-mass model, where the two droplets forming the dimer
oscillate back and forth around their equilibrium positions. 
In order to show this,
an initial small velocity boost is applied to the ground-state 
wavefunctions $\psi _i$, $\psi _i^{ini}=\psi _ie^{\pm ikx}$
where $k=m_i v$. The initial velocity is $v=0.5\times 10^{-10}\,a_0\,E_h/\hbar $. The sign is chosen so that 
the two droplets (1,2) and (2,3) are pushed towards one another at $t=0$.
The system is then left to evolve in real time, by solving the 
time-dependent equations \eqref{GP}.
The time dependence of the bond length $d(t)$ (defined as $d=<x>_{12}-<x>_{23}$,
in terms of the center-of-mass of the two droplets), is calculated and its
Fourier transform $\tilde {f}(\omega)=\int e^{i\omega t}d(t)dt$ provides the dimer vibrational spectrum.
This is shown with solid line in Fig.\ref{fig5}. The peak position in Fig.\ref{fig5} gives
the frequency $\Omega _d/(2\pi) = 576\,Hz $ for the dimer bond-length oscillation.
Similar calculation, done for the case $a_{23}= -70\,a_0$, gives a
lower frequency for the dimer oscillation, $\Omega _d/(2\pi) = 305\,Hz $.
The associated vibrational spectrum is shown with a dotted line in Fig.\ref{fig5}.
Additional features appear in both spectra at lower frequencies 
which might arise from mode-mixing between the dimer bond-stretching motion and
low-energy superfluid (Goldstone) excitations of the 
superfluid background provided by the shared component 2.
Evidence supporting this interpretation will be given when
discussing the vibrational spectra of the diatomic chain in Section IV.

Due to the superfluid nature of the two droplets in the dimer structure,
the hydrodynamic modes describing internal oscillations (multipole modes) 
associated with the superfluid nature of the system
are expected to be excited as well during the dimer dynamics. 
To check for their presence, 
the values of the monopole and quadrupole moments 
$<F(t)>=<\psi _i(t)|F|\psi _i(t)>$ 
are calculated during the dimer oscillations
and later Fourier analyzed, as described in the case of the isolated droplets.

In particular, the monopole and quadrupole spectra of the 
species 1 and the species 3 are computed separately. By assuming that 
during the dynamics the species 2 follows 
the local changes in densities of the other two species in each droplets,
so that the density ratios $\rho _1/\rho _2$ and $\rho _2/\rho _3$ remains close to the 
equilibrium value for the individual droplets, then the computed moments 
should reflect the intrinsic dynamics of the two droplets that make up the 
whole dimer.
Notice that in this way only excitations where the two species in each droplet
move in-phase with respect to each other are considered,
that cause large center-of-mass displacements/bond-length
oscillations with little relative phase change,
and thus emphasizes the mass-spring (bond-stretching) mode.
Excitations characterized by out-of-phase
displacements are therefore not included in the present analysis, but will be
considered in the analysis of the vibrations of the diatomic chain in Section IV.

\begin{figure}[t]
\includegraphics[width=7cm]{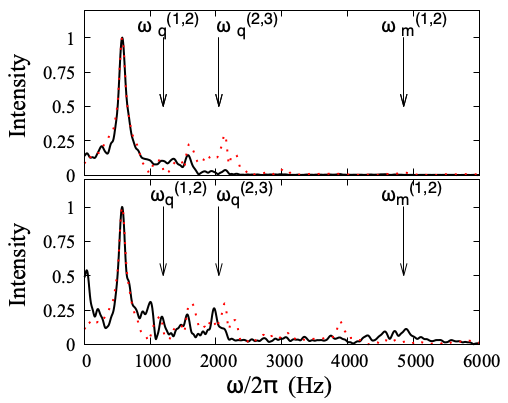}
\caption{Lower panel: spectrum for the monopole oscillations, for the case $a_{23}=-200\,a_0$.
Upper panel: spectrum for the quadrupole oscillations} 
\label{fig6}
\end{figure}

The Fourier transform in the time domain $\tilde {f}(\omega)$ are shown in 
the two panels of Fig.\ref{fig6} for the monopole and quadrupole mode, respectively.
The two curves in each panel show the computed moment (monopole
or quadrupole) in the species 1 (solid line) and the species 3 (dotted line).
Besides the main peak at $576$ Hz coming from the droplet-droplet 
oscillations discussed in Fig.\ref{fig5}, peaks close to the 
monopole and quadrupole frequencies of the individual droplets 
are clearly visible.

Notice that additional small features are present in the Fourier spectrum, especially 
between 1000 and 2000 Hz, revealing a richer excitation
spectrum for the dimer case, where monopole and quadrupole modes are
coupled. Finite-size, nonlinearity effects and the intertwined nature of the 
three species complicate the interpretation of the vibrational spectra, making
a precise assignment of such modes quite difficult.

Similar coexistence between "mass-spring" modes and superfluid (hydrodynamical) modes 
is not new: classical-linear-chain behavior for droplet systems has been observed 
in the case of linear arrays of dipolar droplets.
The vibrational patterns of isolated-droplet crystals that
time-evolve after a small initial kick closely follow the properties of a classical linear chain,
where droplets play the role of atom in classical monoatomic linear chain
\cite{reimann_1}.

\section{IV. The supersolid "diatomic" linear chain}

The dimer configuration described in the previous Section, and its properties,
suggest the possibility of engineering more complex structures 
arranging multiple dimers. The simplest possibility, 
which is studied here, is a linear chain made of
a sequence of droplets ...(1,2)-(2,3)-(1,2)-(2,3)-(1,2)-(2,3)....
I will describe here a simple method, used in the present simulations, 
which could be easily 
translated into an experimental protocol to practically realize such structure.
In the following, the $x$-axis will be the axis of the linear chain of droplets.
Due to the high computational cost of the simulations, 
I will consider in the following only the case with $a_{23}=-200\,a_0$,
although similar phenomenology is expected for the case with $a_{23}=-70\,a_0$ 
(or any intermediate value for that matter).

A {\it uniform} mixture of the three types of atoms is considered, 
subject to an external waveguide potential (the same for all three species)
of the form $(1/2)m_i \omega _{r}^2 r^2$ ($i=1,2,3$) where 
$r^2=y^2+z^2$ is the radial coordinate perpendicular to the
axis of the waveguide, and $\omega _r=(2\pi )\times 300$ Hz.
Due to the use of Periodic Boundary Conditions in the present calculations, 
the system effectively forms an infinitely extended "tubular" geometry
(or, equivalently, a ring geometry if curvature effects can be neglected).
Let $L_x$ be the length of the simulation cell along the tube axis.


An additional weak optical potential is also applied along the $x$-axis, which
has the purpose of modulating, along the tube axis, the uniform mixture 
localized within the waveguide, and trigger the separation into (1,2) and (2,3)
droplets. This potential has the form 
$V(x)=V_0 \,\text{sin}^2(kx)$, where $k=\pi/a_L$ and $a_L$ is the "lattice constant",
i.e. the length of the basic unit cell which will eventually contain a single dimer and
whose periodic repetition will provide the chain sequence ...(1,2)-(2,3)-(1,2)-(2,3)-(1,2)-(2,3)....
I took $V_0=10^{-13}\,E_h$ as a representative value for a weak modulation,
although other choices for this parameter will work as well.

The atom numbers $N_1$, $N_2$ and $N_3$ are chosen so that an integer number of 
dimers like the one shown in Fig.\ref{fig3} 
can be accomodated in a supercell of given length $L_x$ along the x-direction.
Here I took
$N_1=1.4\times 10^5$, $N_2=4.5\times 10^5$ and $N_3=10^5$ atoms,
corresponding to a total of four dimers, i.e. $L_x=4\,a_L$.
A precise choice for the number of atoms or for the simulation cell length $L_x$ is 
however not critical for the
realization of the final chain structure nor for its supersolid character:
even for imbalanced mixtures,
the excess population of one species with respect to the equilibrium value of an isolated droplet
will be expelled from it
and contribute to a uniform background, where the chain modulated structure 
is immersed. In particular, this background is expected to further contribute to the 
chain superfluid fraction (see the following), and therefore to its supersolid properties.
Similarly, a different choice for $L_x$ will results in a compressed/elongated chain, 
without jeopardising the supersolid formation.

During the minimization in imaginary time, the initial homogeneous mixture 
under the effect of the external potentials spontaneously 
separates into (1,2) and (2,3) mixtures, which form a regular 
sequence of droplets of alternating type.
The structure found is an infinitely extended linear chain where the
primitive cell of length $a_L$ contains a "basis" made of the (1,2)-(2,3) dimer.
The external potential (both the radial and the longitudinal ones)
are then removed, and the imaginary-time evolution is continued until convergence.

We note that if the structure minimization is performed 
in the absence of the modulating potential (but still mantaining the radial confining potential),
as a result a highly "defective" structure is eventually obtained,
where the system indeed spontaneously undergo fragmentation in two type
of droplets, but with different sizes of the same type of droplets, and large distortions
of the droplet themselves.

The resulting structure is an equilibrium configuration in vacuum, 
meaning that no confining potential is required to stabilize the final chain structure.
In practice, however, a radial confinement might be kept active
in actual experimental realizations to avoid possible instabilities of the chain due
to axial undulations which may destabilize it.

\begin{figure}[t]
\includegraphics[width=9cm]{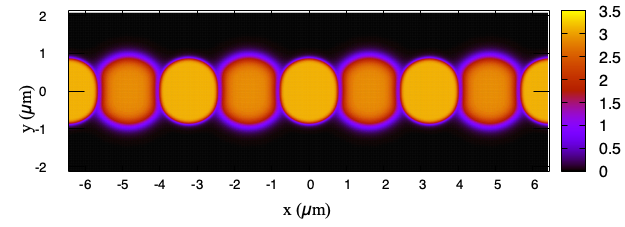}
\caption{Total density maps corresponding to the diatomic chain structure with
$a_{23}=-200\,a_0$.
The density units are the same as in Fig.\ref{fig2}.
} 
\label{fig8}
\end{figure}

\begin{figure}[t]
\includegraphics[width=9cm]{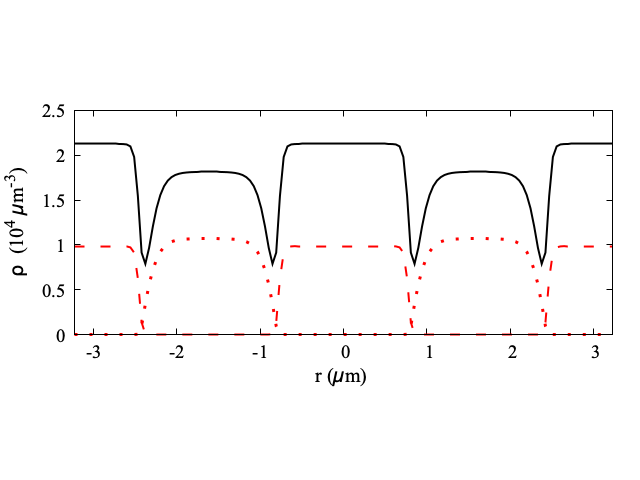}
\caption{Density profiles along the axis of the chain structure 
of Fig.\ref{fig8}.
For clarity, only one half of the chain in Fig.\ref{fig8} is shown here.
Dotted line: species $1$; solid line: species $2$; dashed line: species $3$.  
} 
\label{fig9}
\end{figure}

The final step is to optimize the lattice constant, i.e. 
the parameter $a_L$ is varied (at fixed number of atoms) until the energy is
minimized.
I found in this way for the equilibrium lattice constant the value $a_L=3.23\,\mu m$.
The final ground-state configuration is shown in Fig.\ref{fig8} and Fig.\ref{fig9}.

The bond length (defined as the distance between the center-of-mass of two adjacent droplets) 
turns out to be $d=a_L/2=1.61\,\mu m$, which is 
smaller by $\sim 5\%$ than the bond-length in the isolated dimer in Fig.\ref{fig3} ($d=1.69\,\mu m$),
indicating an increased binding between adjacent droplets when the chain is formed
out of individual dimers. In terms of the mass-and-spring model, this implies
that the force constant $\tilde {K^\prime }$ between adjacent droplets in 
the chain is larger than the one characterizing the oscillations of the isolated dimer,
as shown in the following.

A similar protocol to the one described previously to nucleate 
a diatomic linear chain allows to realize a supersolid diatomic chain in the
case $a_{23}=-70\,a_0$.
The resulting structure (not shown)
has a larger lattice constant, reflecting the lower binding 
between droplets of different types, and it is characterized by a smaller  
overlap between adjacent droplets compared with the one obtained  
with $a_{23}=-200\,a_0$, similarly to the case of the dimer structures.

We must note at this point that during the droplet formation process
in any experimental attempt to create such structure, transient
local excitations (surface modes of nascent droplets,
breathing/vibrational modes, relative-phase distortions between droplets), coming from
the sudden local density rearrangement,
are expected to be created and should be subsequently damped
by redistribution of kinetic energy in the system.
The imaginary-time minimization used here to create the droplet chain
suppresses most of these transients. The effect of excitations
created during droplet formation could be revealed only by a full
real-time dynamics simulation of the whole process, which however
I did not attempt here.

The systems discussed here shows both density modulations and phase coherence
(guaranteed by the common background of the species (2)),
which are the prerequisites for supersolidity.
I have therefore looked for the signature of the free flow of the
superfluid fraction of system through its periodically modulated structure,
namely a finite non-classical translational inertia.

Following Refs.\cite{sep_joss_rica,sepulveda}, the superfluid fraction
$f_s$ can be defined as the fraction of particles that remain at rest in the
comoving frame with a constant velocity $v_x$.
This is found by solving for stationary states the
equation (\ref{GP}) in the
reference frame moving with constant velocity $v_x$, i.e.
\begin{equation}
 i\frac{\partial}{\partial t}\psi _i({\bf r})=
(H_0+i v_x\frac{\partial}{\partial x})\psi _i({\bf r})
\end{equation}
($H_0$ being the Hamiltonian operator within round brackets in Eq.\eqref{GP})
and then computing the superfluid fraction
\begin{equation}
 f_s=1-\lim_{v_x\to 0} \frac{{\big <}(P_{x,1}+P_{x,2}+P_{x,3}){\big >}}{(N_1m_1+N_2m_2+N_3m_3)v_x} \, ,
\end{equation}
where ${\big <}P_{x,j}{\big >}=-i \int (\psi _j ^{\ast} \partial \psi _j/\partial x)\, d\vec r$
is the expectation value of the momentum of the $j$-th species and $(N_1m_1+N_2m_2+N_3m_3)v_x $ is
the total momentum of the system as if all droplets
were moving as rigid bodies.
A non-zero value for $f_s$ reveals global phase
coherence in a periodic system like the one studied here, and therefore supersolid character.




I found in this way, for the chain structure shown in Fig.\ref{fig8}, the value $f_s = 0.71$.
Not surprisingly, almost all the contribution to the superfluid fraction comes
from the shared component 2, which forms a background
with large overlap in the interstitial region between 
one droplet and the next, as it appears from Fig.\ref{fig9}.

Alternatively, one could estimate an upper bound of the 
(component-resolved) superfluid fraction using Leggett's
formula \cite{leggett} separately for each species:

\begin{align}
f^{(i)}_s=(<n _i> <n _i^{-1}>)^{-1}
\end{align}
where 
\begin{equation}
<n _i>=(1/L_x)\int _0^{L_x}dx <\rho _i> 
\end{equation}
\begin{equation}
<n _i^{-1}>=(1/L_x)\int _0^{L_x}dx <\rho _i>^{-1}
\end{equation}
and $<\rho _i>$ is the number density of the ground-state configuration
for the $i$-th species
averaged over the transverse y-z directions. Here $L_x$ is the length of the 
simulation cell along the chain axis.
The calculated fractions are: $f^{(1)}_s=0.05$, $f^{(2)}_s=0.82$, $f^{(3)}_s\sim 0$,
confirming the main role played by the shared component 2.
The system therefore behaves as a composite supersolid
where the supersolid mass transport is dominated by a single
delocalized component (species 2) that connects droplets.


I now discuss some vibrational properties of the supersolid chain structure
described above.

First notice that a spring-and-mass model for the vibrations of an isolated dimer
where the two atoms move back and forth with a periodic motion of frequency $\Omega _d$, 
as discussed in the previous Section, gives
$\Omega _{d}=\sqrt{\tilde {K}/\mu}$, where $\mu =M_{12}M_{23}/(M_{12}+M_{23})$ 
is the reduced mass of the two droplets and $\tilde {K}$ is the force constant.
In the present case we have $M_{12}\sim M_{23}\equiv M$,
and therefore $\Omega _{d}=\sqrt{2\tilde {K}/M}$.
A numerical estimate for $\tilde {K}$ can be obtained by using the computed value of $\Omega _{d}$
from Fig.\ref{fig5}.

The lattice of the supersolids structure investigated here can be described by a
periodically repeated primitive cell containing a "two-atom" basis (the (1,2) and (2,3) droplets), 
and further understanding 
may be obtained by analogy with the dynamics of non-superfluid crystals, i.e.
a classical diatomic linear chain made of particles with alternating masses $M_{12}$ and $M_{23}$,
connected by a spring with constant $\tilde {K^\prime }$, where $\tilde {K^\prime }>\tilde {K}$
(see the discussion following the presentation of Fig.\ref{fig8} and Fig.\ref{fig9}).

\begin{figure}[t]
\includegraphics[width=7cm]{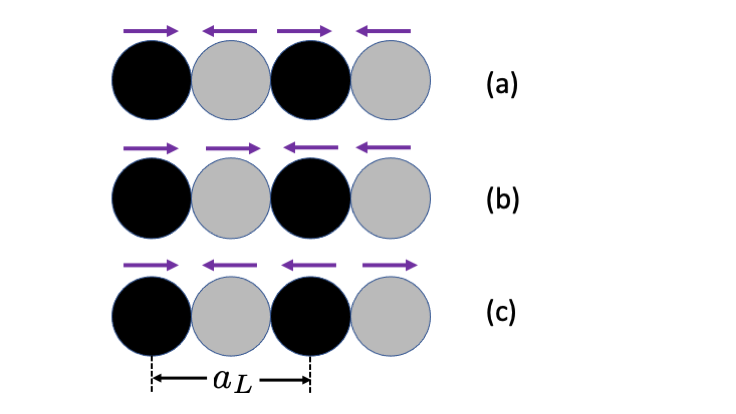}
\caption{Displacement patterns for the different longitudinal modes of a diatomic linear chain:
(a) $k=0$ optical oscillation (with frequency $\omega _O(k=0))$;
(b) $k=\frac {\pi}{a_L}$ acoustic oscillation (with frequency $\omega _A(k=\pi/a_L))$;
(c) $k=\frac {\pi}{a_L}$ optical oscillation (with frequency $\omega _O(k=\pi/a_L))$.
The two vertical dotted lines shows the unit cell of length $a_L$.
} 
\label{fig10}
\end{figure}

In ordinary crystal lattices with two atoms per primitive
cell (of length $a_L$), the phonon branch of a monatomic linear crystal
is replaced by a gapped spectrum,
with the lower acoustic (upper optical) branch being characterized
by in-phase (out-of-phase) motions between nearest
neighbors.
Fig.\ref{fig10} shows schematically the displacement patterns 
of atoms in a diatomic linear chain at two relevant points in the 
1$^{st}$ Brilloun Zone (BZ) along the chain axis, i.e. the zone-center $k=0$
and the point at the zone boundary, $k=\pi/a_L$.
The longitudinal acoustic mode at $k=0$ (not shown in the Figure) 
is just a rigid translation of the chain as a whole 
along the x-axis, and as such has zero frequency.

The frequencies of the longitudinal modes in classical diatomic linear chain, whose patterns are shown in
Fig.\ref{fig10}, are 
$\omega _O(k=\pi/a_L, \text{optical})=\sqrt{2\tilde {K}/M}= \Omega_d$,
$\omega _A(k=\pi/a_L, \text{acoustic})=\sqrt{2\tilde {K}/M}= \Omega_d$,
$\omega _O(k=0, \text{optical})=\sqrt{2\tilde {K}/\mu}= \sqrt{2}\Omega_d$.
The two zone-boundary modes are degenerate because the masses $M_{12}$ and $M_{23}$
of the two droplets are almost equal, as already discussed.

In general, for linear supersolids a Goldstone branch of superfluid
phonons associated with the condensate order parameter 
is expected to appear, besides the crystal phonon branch
characteristic of a regular solid \cite{pita16}.
The periodic modulation of the density reduces
the superfluid fraction \cite{leggett} and
the lower phonon branch (superfluid mode) has a
second-sound character, while the upper branch (lattice phonon mode) consists
of first-sound-like modes \cite{sind24}.

In fact, recent experimental observation \cite{tanzi1} in dipolar quantum gas
showed that when the system enters supersolid state, low energy compressional modes 
(which emerge naturally from the hydrodynamics equations of superfluidity and
must therefore be associated with the finite superfluid fraction) bifurcates
into two distinct excitations 
(similarly with the appearance of the gapless Goldstone excitations expected for 
homogeneous supersolid): the higher-frequency mode is related to the lattice deformations
(relative displacements of the droplets making the chain), while the lower mode is instead related to
compressional excitations of the superfluid components.

In order to excite the proper (longitudinal) oscillation mode in the supersolid chain 
described here I proceeded as follows:
a small perturbation in the form of an axial
density modulation along the supersolid axis is 
added to the equilibrium density at t=0:

\begin{equation}
\rho (\vec r)=\rho _0 (\vec r)[1+\epsilon \,\text{cos}(k x)]
\end{equation}
with wave-vector $k$. 
Here $\rho _0(\vec r)$ be the
total ground-state density and $\epsilon $ is a small number.
In order to reduce the computational cost 
calculations have been done for a linear chain which has a length along the x-axis
which is one half of that illustrated in Fig.\ref{fig8}, i.e. the basic cell 
for real-time dynamics is composed of two unit cells, each of length $a_L$ containing two
droplets (1,2) and (2,3). 
I considered two values for $k$, i.e. $k=0$ (center of the 1$^{st}$
Brillouin zone along the chain axis), and $k=\pi/a_L$ (surface of the 1$^{st}$
Brillouin zone).
The amplitude $\epsilon $ of the perturbation must be small: here I took $\epsilon =0.025$.
The system is thus left to evolve in real time.
The density $\rho (\vec r,t)$ is computed at regular time intervals, and
from the computed values the projection of the density onto the 
$k$-mode is calculated:

\begin{figure}[t]
\includegraphics[width=7cm]{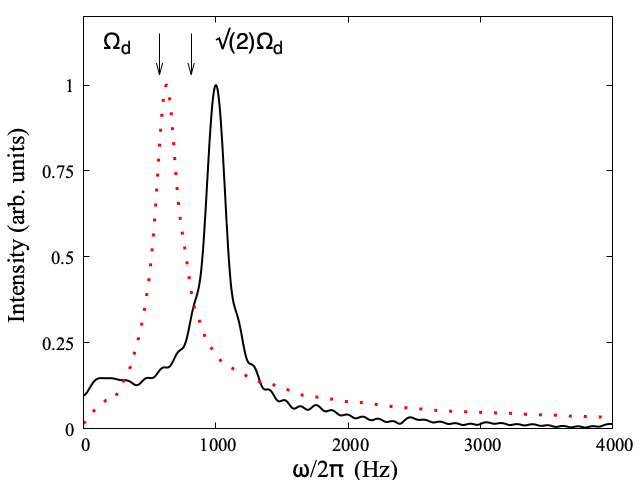}
\caption{Spectrum at $k=0$ (solid line) and $k=\pi /a_L$ (dotted line).
The vertical arrows indicate the positions expected for the classical diatomic
chain modes.
} 
\label{fig11}
\end{figure}

\begin{equation}
\rho (k,t)=\int _0 ^{L_x}dx \,e^{-ikx}\int dy \int dz \rho (x,y,z,t)
\end{equation}

Finally, $\rho (k,t)$ is Fourier transformed into the frequency domain: 

\begin{equation}
\rho (k,\omega )=\int dt \,\rho (k,t) \,e^{i\omega t}
\end{equation}
to provide the mode frequencies as peaks in $\rho (k,\omega )$
(a similar method is used in Ref.\cite{platt} to probe one-dimensional supersolid states,
but where a
periodic optical potential is applied to excite the supersolid and determine its 
density excitation frequencies).

The above approach is feasible for any finite $k$-value whose wavelength $2\pi/k$ fits
the cell length along the x-direction.
In order to compute the optical modes at $k=0$ another method is therefore required.
I used here a simple strategy where a kick to the droplets composing the chain is applied to excite the 
desired displacement pattern associated with the mode under study.
An initial small velocity boost is applied to the ground-state 
wavefunctions $\psi _i$, $\psi _i^{ini}=\psi _ie^{\pm ikx}$
where $k=m_i v$. The initial velocity is $v=0.5\times 10^{-10}\,a_0\,E_h/\hbar $. The sign is chosen so that 
the droplets (1,2) and (2,3) initial displacements match the pattern
characterizing the optical mode at $k=0$ (case (a) in Fig.\ref{fig10}).
The system is thus left to evolve in real time and the
droplets relative displacements are Fourier analyzed in order to find the 
associated frequencies.

The results are shown in Fig.\ref{fig11} for both cases, $k=\pi/a_L$
and $k=0$.
Two main peaks are clearly visible: the lower frequency one ($k=\pi/a_L$) 
is associated with the (degenerate) acoustic and optical modes 
at the BZ boundary (cases (b) and (c) in Fig.\ref{fig10}), while the higher 
frequency mode ($k=0$) is associated with the zone-center optical mode (case (a) in Fig.\ref{fig10}).
Both frequencies are higher than those expected using the force constant 
estimated from the isolated dimer vibrations (shown by vertical arrows in the Figure).
This is expected
since the force constant for the chain is larger than the 
one for the dimer due to the increased binding in the chain with respect 
to the dimer case, as discussed before.
We finally mention that 
dispersion relations showing similarities with a linear diatomic crystal
has been found recently \cite{kir24} in binary dipolar supersolid.

The dynamics described previously 
restricts the analysis to in-phase motion of the components in each droplet, and 
therefore emphasizes the mass-spring (bond-stretching) mode discussed above. 
To check whether low-energy, out-of-phase (odd-parity) modes - such as Josephson or Goldstone 
type phase oscillations that are connected to global phase coherence in supersolids \cite{biagioni,guo} - are present, 
we performed additional real-time simulations where an initial small relative phase difference 
with $k=\pi/a_L$ (i.e. at the boundaries of the first BZ) 
was imprinted on the shared $^{39}$K component (which is
the relevant one for the superfluid character throughout the system).
In particular, the initial wavefunction
$\psi _2(\vec r)$ is obtained from the ground-state one $\psi _2^{(0)}(\vec r)$
as
\begin{equation}
\psi _2(\vec r,t=0)=\psi _2^{(0)}(\vec r) e^{i\Delta \phi cos(\pi x/a_L)}
\end{equation}
where the amplitude $\Delta \phi $ is chosen small ($\sim 0.02$) to remain in the linear regime.
Then, during the real time evolution, I computed

\begin{equation}
\psi _2(k,t)=\int _0 ^{L_x}dx \,e^{-ikx}\int dy \int dz \psi_2(x,y,z,t)
\end{equation}

Finally, $\psi _2 (k,t)$ is Fourier transformed into the frequency domain: 

\begin{equation}
\psi_2 (k,\omega )=\int dt \,\psi_2 (k,t) \,e^{i\omega t}
\end{equation}
to provide the mode frequencies as peaks in $|\psi_2 (k,\omega )|$.
The resulting spectrum is shown in Fig.11, where $|\psi_2 (k=\pi/a_L,\omega ) |^2$ is plotted, 
showing low-frequency branches (below 500 Hz)
which should be associated with out-of-phase oscillations of the condensate phase (Goldstone/Josephson character) 
distinct from the higher-frequency lattice phonon branches
(a more detailed analysis of the shown spectrum is however difficult, for the same reasons 
described when commenting the spectra in Fig.\ref{fig6}).
These results provide independent dynamical evidence of global phase 
coherence and further support the supersolid nature of the diatomic chain structure.

\begin{figure}[t]
\includegraphics[width=7cm]{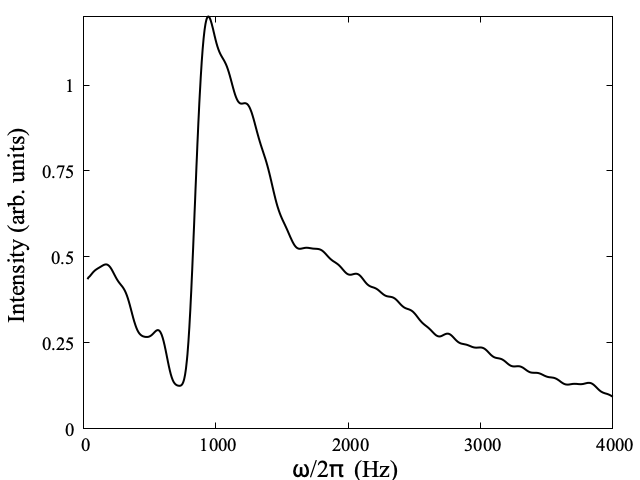}
\caption{Spectral function $|\psi_2 (k,\omega)|$ at $k=\pi /a_L$. 
} 
\label{fig12}
\end{figure}

\section{V. Conclusions and outlook}

In this work, I have theoretically investigated the formation and properties 
of a supersolid structure in a three-component ultracold Bose gas 
mixture composed of $^{23}$Na, $^{39}$K, and $^{41}$K atoms, 
in the form of an extended linear chain structure, stable in vacuum.
This structure is made by periodic repetition of the basic building block represented
by a "dimer" unit consisting of two weakly bound quantum droplets, 
made of (1,2) and (2,3) species.

The dynamics of an isolated dimer shows the coexistence of 
stretching vibrations as in mass-and-spring model, where the dimer bond length 
oscillates harmonically in time, and hydrodynamical modes 
associated with the superfluid nature of the system.

A protocol is presented, which could be implemented in 
experimental realization, to create a linear chain structure
starting from a homogeneous three-component mixture subject 
to a suitable external potential, which results in the 
spontaneous formation of an alternating sequence 
of ($^{23}$Na,$^{39}$K) and ($^{39}$K,$^{41}$K) droplets.

This "diatomic linear chain" structure exhibits both periodic density modulations arising 
from the droplet ordering and global phase coherence, 
facilitated by the shared $^{39}$K component, therefore satisfying the criteria for supersolidity.
A relevant superfluid fraction is indeed found by
computing the non-classical translational inertia of the system.
The excitation spectrum of the supersolid chain is
investigated by probing its response to density perturbations.
I identified longitudinal vibrational modes corresponding to 
crystal lattice oscillations like acoustic and optical phonon modes 
expected from a classical diatomic chain, coexisting with 
low-frequency out-of-phase oscillation of the condensate phase.

The supersolid phase studied here is qualitatively 
distinct from previously studied realizations,
the broken translational symmetry and global phase coherence arising
from a mediated droplet binding mechanism rather than from long-range forces or
momentum-space interference.
In contrast to dipolar systems (where long-range dipole-dipole
interactions and roton softening set the periodicity of the supersolid)
this provides several new handles (interspecies scattering lengths, relative populations and mass
imbalance) to tune the lattice spacing, the coupling strength between droplets, the
relative sizes of the two types of droplets and the resulting superfluid fraction,
potentially accessing regimes that are difficult to reach in dipolar or spin-orbit platforms.
Therefore this supersolid chain represents a
novel tunable quantum phase in multi-component systems,
with potential for experimental realizations using state-of-the-art ultracold atom setups.
Although the experimental implementation of a three-species setup is likely
more demanding than two-component or single-species realizations,
the extra experimental complexity might be compensated by the increased
tunability and by the qualitatively new phenomena this platform can explore.

Future research directions could include exploring the stability and dynamics 
of the supersolid chain under various external perturbations, including the effect of transverse 
excitation modes,
as well as investigating the effects of dimensionality and geometrical 
constraints on the formation and properties of such structure.
Another interesting line in future research would be the study of
vortices hosted by crystal configurations of the three-component supersolid, which is
the most direct evidence of superfluidity in a 
supersolid structure response to rotation.
For instance, a 2-D periodic array of (1,2) and (2,3)
droplets subject to rotation around the axis perpendicular 
to the plane of the 2D "crystal", would be the ideal playground to study
the formation and stability of quantized vortices \cite{vort}.

\bigskip

\acknowledgments
I wish to thank G. Mistura and L. Salasnich for useful conversations.
This work is supported by the Italian MIUR under the PRIN2022 Project No. 20227JNCWW.


\begin{thebibliography}{99}
\bibitem{note}
For the numerical evaluation of Eq.~(\ref{E_qf})
it is convenient to use the following transformation 
\begin{equation}
\int _0 ^\infty g(k)dk=\int_0 ^{\pi/2}g[\tan(t)]\,\frac{dt}{\cos^2(t)} 
\end{equation}
The right-hand side integral has been computed numerically using a
Second Euler-McLaurin summation formula refined until some specified 
degree of accuracy is achieved~\cite{Pre99}.
\end{thebibliography}
\end{document}